\documentclass[a4paper,twoside,10pt]{article}

\usepackage[USenglish]{babel} 
\usepackage[T1]{fontenc}
\usepackage[ansinew]{inputenc}
\usepackage{commath}

\usepackage{lmodern} 

\usepackage{graphicx} 

\usepackage{caption}
\usepackage{subcaption}

\usepackage{amsmath}
\usepackage{amsthm}
\usepackage{amsfonts}

\usepackage{fancyhdr} 

\usepackage[font={small,it}]{caption}

\usepackage[amssymb]{SIunits}
\usepackage{color}

\newcommand{\atan}[2]{\mathrm{atan2}\left({#1},{#2}\right)}

\newcommand{\twoD}{{2\mathrm{D}}}
\newcommand{\threeD}{{3\mathrm{D}}}

\begin{document}

\pagestyle{empty} 

\title{Direct Wing Design and Inverse Airfoil Identification with the Nonlinear Weissinger Method}
\author{Maximilian Ranneberg\thanks{m.ranneberg@tu-berlin.de}}
\date{} 


\maketitle

\pagestyle{plain} 
\begin{abstract}
A vortex-lattice method for wing aerodynamics that uses nonlinear airfoil data is presented.
Two applications of this procedure are presented: Direct Design of a Flying Wing and Inverse Identification from wind tunnel measurements with low-aspect ratio wings.
A Newton method is employed, which not only allows very fast solutions to the nonlinear equations but enables the calculation of static and dynamic stability and control derivatives without further cost.
\end{abstract}
\section{Introduction}
Fast methods to obtain aerodynamic characteristics of aircraft are and have been of great interest in all aircraft disciplines.
The governing Navier-Stokes equations result in far too complex calculations during the initial phases of aircraft design, and for small projects are infeasible even for post-design and evaluation.
For the purpose of identifying the aerodynamic characteristics of aircraft, the tools used today are still based on the theoretical framework of the 40s.
These are the Vortex-Step, Vortex-Lattice and Panel methods.
They are all based on the same idea: Assuming non-rotational, incompressible flow and discretization this flow by combinations of fundamental solutions at certain places on the wing and body.
The differences in these methods lies solely on the amount, and position, of these solutions.
This results in surprisingly accurate predictions even for simple discretizations used in the 40s.
Today, Panel Methods are among the most widely used aircraft design tools.

The drawback of these methods is simple.
For Vortex-Step and Vortex-Lattice methods, no knowledge whatsoever is used on the shape of the airfoil, besides partly camber which effectively results in a change of zero-lift angle of attack.
This results in significant errors for configurations with unconventional airfoils, for example high-lift configurations and low-speed small aircraft.
No stalling characteristics are used, and no profile drag can be estimated.
Even more detailed Euler Equation solvers cannot properly model the drag, although some promising developments are being made \cite{3DDrela,Cart3D}.
These approaches are still quite experimental though and far from being used in common applications.

However, two-dimensional data of profile lift, drag and moment are available.
Reliable simulation of 2D profiles has been feasible for years, detailed measurements are available and today, full Navier-Stokes solutions of the 2D problem are possible within hours on a common machine.
The software XFLR\footnote{http://www.xflr5.com}, if panel or vortex-lattice method is chosen, uses the 2D data calculated from XFOIL\cite{XFOIL} to append viscous drag related to the current local lift and Reynolds-Number and to estimate the maximum local lift coefficient.

Methods that try to use the 2D data in a more involved way exist since the 40s.
In \cite{Nonlinear47} a method of incorporating the local lift coefficient into a lifting-line method is proposed.
There, the nonlinear equations for the vortex strength along the span is solved using fixed-point iteration.
This methodology is, for example, implemented in the software \emph{miarex}, which is now part of XFLR if the method is chosen.
The drawback are the drawbacks of the lifting-line method: For swept and low aspect ratio wings the results should be handled with caution.

Another method with a similar approach was proposed in \cite{Nonlinear76}.
It uses the 2D data together with a generalized Vortex-Step or Weissinger method, which takes into account sweep and low aspect ratio.
It is based on the generalization of the tangential-flow condition commonly applied to Vortex-Lattice Methods which will be discussed later.
There, too, a fixed-point iteration is used for the resulting equations.
In \cite{NonlinearStability} the method has been applied to obtain simplified nonlinear dynamics equations for aircraft for control analysis.
More recent studies, for example in \cite{VanDam} a similar technique is used to obtain corrections to the classic linear Vortex-Lattice Methods.
Instead of using a different boundary condition, an ad-hoc change in the local angle of attack is assumed.
However, one could argue that the generalized boundary conditions of \cite{Nonlinear76} lead to a similar approach.
These conditions effectively modify the classic local downwash with another downwash term, resulting in a local change in angle of attack (albeit slightly more motivated by theory and dependent on the local vortex strength).
Another approach been presented in \cite{Decambering}, which gives an excellent overview of the results thus far and where the change in local angle of attack is interpreted as a change in camber due to the flow separating from the wing earlier.
There, the local lift and moment data is used for changing a vortex-lattice method with two cord-wise panels with a two-parameter decambering function and the resulting nonlinear equations are solved using a Newton method.
Camber of an airfoil describes the asymmetry of an airfoil and results in non-zero lift at zero angle of attack.
Decambering can be interpreted as a means to incorporate the increase in boundary layer thickness towards the trailing edge and the effective reduction of lift due to the reduction of (positive) camber into a three-dimensional method \cite{Decambering2}.
Subsequent applications of this work have been published, for example in \cite{exDecambering}, where an approximation of the method is used for simulation and control.
However, all methods share the same goal: Obtain a solution at every section, such that the global lift due to vortex strength and the resulting effective angle of attack is in accordance to the lift coefficient due to this angle of attack.

Here, a method is described based on the principles derived in \cite{Nonlinear76}, which is similar to the method described in \cite{NonlinearStability}.
The main reason for this basis is the elegance of generalizing the boundary conditions beyond linear airfoils.
The boundary conditions in \cite{Nonlinear76} will be motivated by revisiting the $\frac{3}{4}$ cord theorem and the relation to the generalized boundary conditions.
In contrast to the method employed in \cite{Decambering}, the generalized boundary conditions do not lead to the generalization of two cord-wise parameter variations and thus only to a single cord-wise lattice discretization.

A Newton method is employed and results in fast results with usually only one or two iterations during an angle sweep.
The results are used for detailed aerodynamic analysis for static and dynamic coefficients.
Derivations of these coefficients are possible that do not necessitate a solution to the nonlinear equations, but can be obtained using the gradient of the equations.
Since the gradient needs to be evaluated for the Newton method anyway, the dynamic derivatives are a convenient byproduct of the method.

Additionally, the method is used in an inverse fashion to correct windtunnel results obtained with wings.

\section{The Solution Method}
The Vortex-Lattice method relies on the discretization of the solution to the incompressible Euler equations.
Any closed path of constant vortex-strength is a solution to the equations, and the closed paths used here are classic Horseshoe paths with a tip on the quarter cord of the local wing surface, two legs along the cord, two legs from the trailing edge aligned with the free-stream airspeed and a connecting line at infinity, which results in no downwash.
First, the basics of these methods is explained, which are the Biot-Savart Law for the downwash, the Kutta-Joukowski Law for the lift and the Pistolesi Theorem for the boundary conditions.
Special attention is necessary during the formulation of the boundary conditions, since they are modified to enable nonlinear airfoil data.
In the remaining section, it will always be assumed that $v_a$ is the unit vector of the airspeed direction. 
That is, $\abs{v_a}=1$.

\subsection{Downwash}
The downwash induced at some point $m$ from a vortex path $S$ of constant vortex strength $\Gamma$ is given by the Biot-Savart law
\begin{equation}
D_m = \frac{\Gamma}{4\pi}\int_S \frac{\mathrm{d}l\times(r(s)-m)}{\left|(r(s)-m)\right|^3}.
\end{equation}
If the path $S$ is a linear path between two points $r_0,r_1$ the downwash has a simple analytic anti-derivative, given by
\begin{equation}
D_m^{r_1, r_2} = \frac{\Gamma}{4\pi} (r_1-r_0)^T\left(\frac{r_1-m}{|r_1-m|} - \frac{r_0-m}{|r_0-m|} \right) \frac{(r_0-m)\times(r_1-r_0)}{|(r_0-m)\times(r_1-r_0)|^2}.
\end{equation}
All parts of the path given by a Horseshoe vortex can be calculated.
Points at an infinite distance can be calculated, too, by using the limit as $r\to\infty$.
For example, the trailing leg in the airspeed direction from the trailing edge $r_t$ can be written with $r_0=r_t$, $r_1=r_t+s v_a$ and with the assumption that $\abs{v_a}=1$ as
\begin{align*}
D_m^{r_t, \infty} &= \lim_{s\to\infty} \frac{\Gamma}{4\pi} (s v_a^T\left(\frac{r_t+s v_a-m}{|r_t+s v_a-m|} - \frac{r_t-m}{|r_t-m|} \right) \frac{(r_t-m)\times(s v_a)}{|(r_t-m)\times(s v_a)|^2}\\
&= \lim_{s\to\infty} \frac{\Gamma}{4\pi} (v_a^T\left(\frac{r_t+s v_a-m}{|r_t+s v_a-m|} - \frac{r_t-m}{|r_t-m|} \right) \frac{(r_t-m)\times v_a}{|(r_t-m)\times v_a|^2}\\
&= \frac{\Gamma}{4\pi} \left(1 - v_a^T \frac{r_t-m}{|r_t-m|} \right) \frac{(r_t-m)\times v_a}{|(r_t-m)\times v_a|^2}.
\end{align*}
Another special case which will be discussed during Pistolesi's Theorem is $r_0=-s v$, $r_1=s v$ which results in
\begin{align}
D_m^{-\infty, \infty} &= \lim_{s\to\infty}  \frac{\Gamma}{4\pi} (2 v)^T\left(\frac{sv-m}{|sv-m|} - \frac{-sv-m}{|-sv-m|} \right) \frac{(-sv-m)\times(2v)}{|(-sv-m)\times(2v)|^2}\\
&=  -\frac{\Gamma}{2\pi} \frac{m\times v}{|m\times v|^2} \label{eq:pistolesiHelp}
\end{align}
All of these downwash velocities are vectors.
The downwash in the direction of interest, usually downwards $(0,0,-1)^T$ with respect to the local surface axis, can be evaluated by using the scalar product with the direction of interest.

\subsection{Boundary Conditions}
With the discretized Vortex paths the possible solutions are constrained to all linear combinations of the individual $\Gamma_i$.
The lift associated with the vortex combination is given by the Kutta-Joukowski law, which states
\begin{equation}
L = v_a\times\Gamma.
\end{equation}
The values of $\Gamma_i$ are found by defining sufficient boundary conditions.
The classic Weissinger method uses the tangential flow condition at the $\frac{3}{4}$cord.
At this point, the three-dimensional downwash $D_{\frac{3}{4}}$ induced by all elements should be equal to the geometric angle of attack, resulting in an effective angle of attack of zero and a flow tangential to the lifting surface.
However, there are more subtleties associated with this boundary condition.

The theorem of Pistolesi states that at the $\frac{3}{4}$cord the angle of attack results in the correct lift associated with the vorticity.
A 2D airfoil, or infinitely long wing, with a constant sweep $\gamma$ is considered.
The lift slope of such a wing is given by $c_l^{\twoD}(\alpha) = 2\pi\cos^2\gamma\alpha$ as derived from simple swept wing theory.
Which, basically, reduces the effective airspeed due to the normal direction with respect to the nose by $\cos\gamma$. 
Since the lift depends quadratically on the airspeed lift is reduced by $\cos^2\gamma$. 
At the $\frac{1}{4}$cord a single vortex path with strength $\Gamma$ runs along the wingspan to infinity.
The lift coefficient of this wing with this vortex strength is given by $v_a \times \Gamma = \cos\gamma\Gamma$.
The downwash at the $\frac{3}{4}$ line is given by
\begin{align}
u &= (\sin\gamma,\cos\gamma,0)^T,\ 
\mathrm{d}l = u \mathrm{d}y,\ 
h = \left(\frac{1}{2},0,0\right)^T,\\
D_{\frac{3}{4}}^{\twoD} &= \frac{\Gamma}{4\pi}\int_{-\infty}^\infty \frac{\mathrm{d}l \times (yu-h)}{|yu-h|^3}\\
&\stackrel{\eqref{eq:pistolesiHelp}}{=}\frac{\Gamma}{2\pi\cos\gamma} 
\end{align}
Thus, the downwash results in the angle of attack associated with the vortex strength.

The boundary condition of the classic Weissinger method enforces this condition for the three dimensional downwash.
Note that the tangential-flow condition uses the sweep implicitly, since the assumption is that the terms cancel each other out.
The boundary condition of the Weissinger method can be written in a different way that does not use the assumption of a two-dimensional lift slope of $2\pi\cos^2\gamma$ but instead uses a general two-dimensional lift:
\begin{equation}\label{eq:BasicNonlinearEquation}
c_l^{\twoD}(\alpha-D_{\frac{3}{4}}^{\threeD}\Gamma + D_{\frac{3}{4}}^{\twoD}\Gamma) = \abs{v_a \times \Gamma}
\end{equation}
That is, an effective local angle of attack is used to search for a solution $\Gamma$ where the effective local angle of attack results in local lift coefficients equal to the lift given by $\Gamma$.
For a linear lift slope, the equation results in the tangent-flow conditions as already shown in \cite{Nonlinear76}, since the 2D downwash times the lift slope equals the term $\abs{v_a \times \Gamma}$ as shown above.
It should be noted that the tangent-flow conditions are in fact a misnomer. 
For all other positions on the wing, for example the trailing edge, there is no tangential flow.
It just happens to fit the interpretation of intuitive boundary conditions, if the $\frac{3}{4}$cord control point is chosen.

\subsection{Nonlinear Coupling}
In the previous section the boundary condition of the Weissinger method were reformulated without explicitly stating the 2D lift slope.
These generalized boundary conditions \eqref{eq:BasicNonlinearEquation} can be used with different two-dimensional lift slopes.
And since nonlinear equations are used, it seems appropriate to use the following equation
\begin{equation}\label{eq:NonlinearEquation}
c_l^{\twoD}(\alpha-\mathrm{atan}(D_{\frac{3}{4}}^{\threeD}\Gamma + D_{\frac{3}{4}}^{\twoD}\Gamma)) = v_a \times \Gamma,
\end{equation}
since the induced downwash is in fact not an angle, but the ratio of induced velocity and free-stream velocity which is similar only for small downwash velocities w.r.t. the free-stream velocity.
A damped Newton method is employed to solve the nonlinear equations.



\subsection{Induced Drag}
The induced drag is calculated by using the common Trefftz-Plane Analysis far behind the wing. There, parallel to the trailing edge, the downwash is calculated by the same formula derived from the Biot-Savart law, but special care is necessary for the limit as $r_0\rightarrow\infty$.
Here, the control-point $m$ lies in the Trefftz-Plane and the only downwash contributions are the trailing legs.

\subsection{Washout and Angle of Attack}
In the implementation, the angle of attack is derived completely from the given geometry and the current airspeed vector.
The washout of the airfoil sections is directly applied to the cord-wise legs of the vortices.
This leads to non-symmetric legs.
An example of the geometry is given in Fig.\ref{fig:wing}, where the $\frac{1}{4}$cord line is given but the control points are not shown.

\subsection{Geometric Angle of Attack}
For simple Horseshoe-Vortex geometries with legs and assumed airflow parallel to the x-axis, the geometric angle of attack is given by the angle of attack of the airflow and additionally the angle of attack resulting from the washout. 
More complex geometries with rotations and non-parallel legs result in less trivial geometric angles of attack. 
The surface is given by the vectors $\hat\theta = u_{\Gamma_{yz}}$ and the cord direction ${\hat c}$. 
The local geometric angle of attack, perpendicular to this surface, can be calculated as follows:
The goal is to find an angle $\alpha_0$ about which the cord direction is rotated to result in a maximal scalar product with the airspeed $v_a$.
\begin{align}
\alpha_0 &= \mathrm{max}_\alpha (\mathrm{Rot}_{\hat\theta}(\alpha) c)^T v_a\\
&= \mathrm{max}_\alpha ({\hat c} \cos\alpha + \sin\alpha \hat\theta \times \hat c + (1-\cos\alpha) {\hat c}^T \hat\theta {\hat\theta})^T v_a \\
\Rightarrow 0 &= ((-{\hat c} +{\hat c}^T \hat\theta {\hat\theta})\sin\alpha_0 + (\hat\theta \times \hat c)\cos\alpha_0)^T v_a \\
\Rightarrow \alpha0 &= -\mathrm{atan}\left(\frac{(\hat\theta \times \hat c)^T v_a}{(-{\hat c} +{\hat c}^T \hat\theta {\hat\theta})^T v_a}\right)
\end{align}
The denominator within the atan function is only zero, if and only if the airspeed is completely perpendicular to the cord-wise direction.

\subsection{Sideslip}
The influence of sideslip is incorporated directly by the direction of $v_a$ and results in a different three-dimensional downwash due to the trailing wake as well as in the direction of the local lift.
The geometry is unchanged, the bound vortex and the legs on the wing surface are still in sweep-wise direction or cord-wise direction, respectively.
This approach of calculating the influence of sideslip has been originally proposed by Weissinger\cite{Weissinger} and is in widespread use today, for example in the software XFLR.
It differs, however, from the approach in \cite{Nonlinear76} where sideslip is modeled using asymmetric sweep angles.

\subsection{Stationary Coefficients}
The stationary coefficients can be deduced directly from the solution of the Vortex strength. 
With the generalized Kutta-Joukowski law, the forces due to lift can be calculated, and the drag is given as well.
Assuming that all forces act on the $\frac{1}{4}$cord point, the aerodynamic moments are given as well.
Here, the airfoil moment is added to the total moment.

\subsection{Dynamic Coefficients}
The dynamic linear coefficients can be calculated without reevaluating the downwash or search for a new $\Gamma$.
Exemplary the roll-damping coefficient at some angle of attack with zero sideslip is derived.
The roll damping coefficient is given w.r.t. the non-dimensional roll-rate $p = \frac{p_{real} \mathrm{mac}}{v_\infty}$.
If the stationary local velocity is $v_0 = v_\infty \hat{v_a}$ with $\hat{v_a}=(\cos\alpha,0,\sin\alpha)^T$, the new velocity at a distance $r$ from the reference point or rotation is given by $v = v_0 + \omega_{real} \times r$.
For rolling only, this is $v = v_0 + p_{real} (0,-r_z,r_y)$.
The new angle of attack and sideslip can be linearized with the atan2 derivative and is given by
\begin{align}
	\alpha' &= \atan{v_z}{v_x} = \atan{\sin\alpha + \frac{r_y}{h} p }{\cos\alpha} \\
	&\approx \alpha - \cos\alpha \frac{r_y}{h} p\\
	\beta' &= \atan{v_y}{v_x} = \atan{-\frac{r_z}{h} p }{\cos\alpha} \\
	&\approx \beta + \cos\alpha \frac{r_z}{h} p
\end{align}
These changes in the angle of attack result in a local change of the angle of incidence $\alpha_i$ and in a change of direction of the lift, but also in a slight change of the vorticity.
Equation \eqref{eq:NonlinearEquation} can be approximated around the current $\Gamma$ and $\alpha$ and solved for the changes $\Delta\Gamma$ due to $\Delta\alpha$. 
With the current local effective angle of incidence $\alpha_{eff}$, the complete downwash operator $D$ and the local lift slope $c_l'(\alpha_{eff})$ the change in vortex strength is given by the linear equation
\begin{equation}\label{eq:linearDynamicEq}
\Delta\alpha_i = \left(\frac{\hat{v_a} \times \hat\Gamma}{c_l'(\alpha_{eff})} + \frac{D}{1+\alpha_{eff}^2}\right) \Delta\Gamma
\end{equation}
With the new vortex strength and the new lift and drag directions, all forces and torques can be reevaluated by the Kutta-Joukowski Law and the induced drag definition.
Finite differences can then be used to calculate the coefficient.
That is, 
\begin{equation} 
L_p = \frac{L(\alpha,\Gamma) + L(\alpha+\Delta\alpha_p,\Gamma+\Delta\Gamma_p)}{p}.
\end{equation}
The linear equation \eqref{eq:linearDynamicEq} can be used for all dynamic derivatives, not only for the rotational rates but also for the aircraft velocities and linear $\alpha$,$\beta$ derivatives.
\newpage
\begin{figure}
	\centering
		\includegraphics[width=0.99\textwidth]{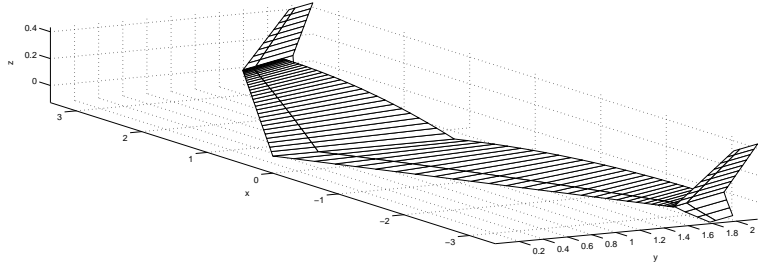}
	\caption{Geometry of a wing. The washout is directly included in the vortex geometry. The $\frac{1}{4}$cord line is hinted and defines the bound vortex position.}
	\label{fig:wing}
\end{figure}
\section{Results for a Flying Wing}
\begin{figure*}[tb]
	\centering
		\begin{subfigure}[b]{0.49\textwidth}
		\includegraphics[width=\textwidth]{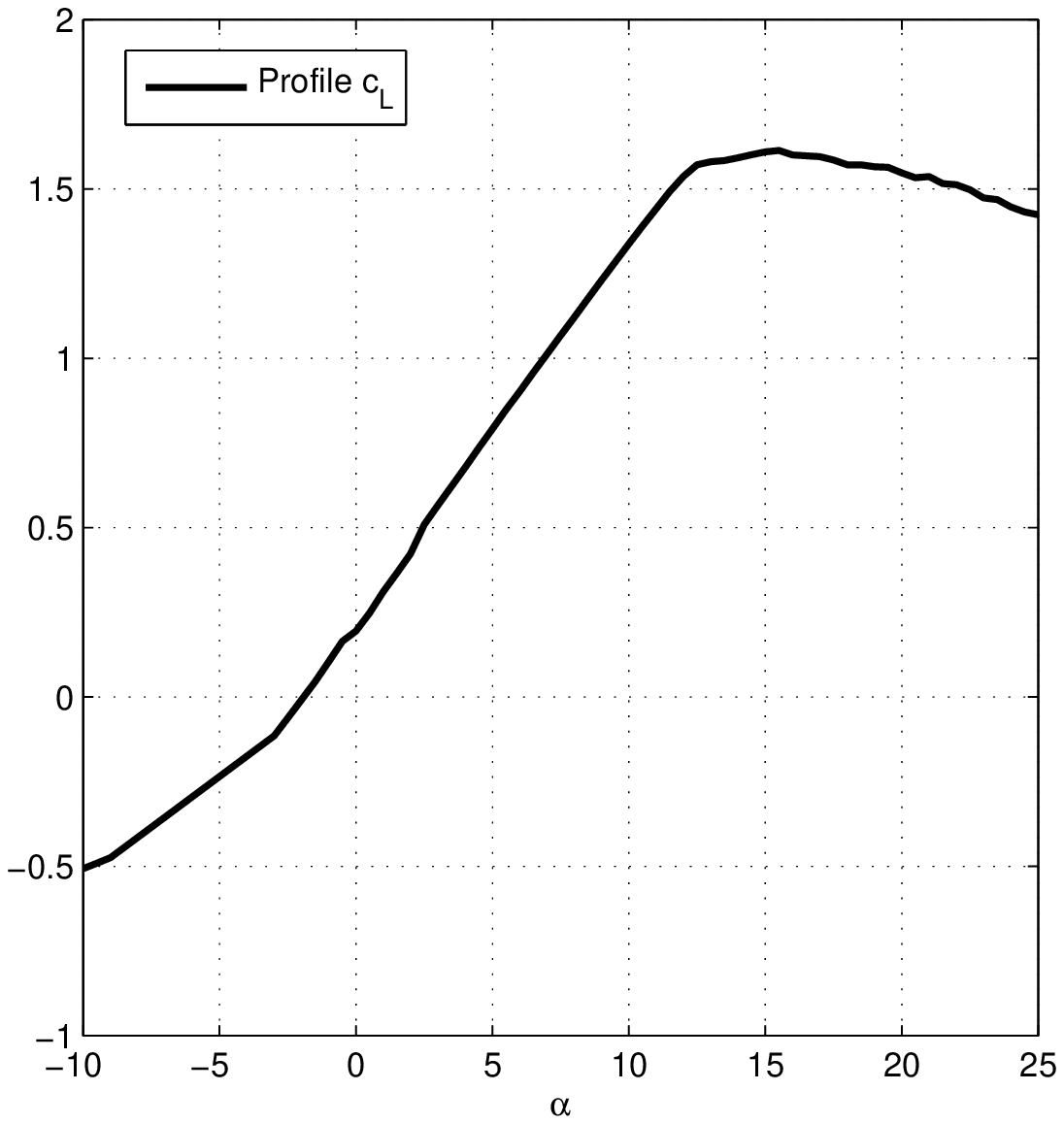}
		\label{fig:profLift}
		\end{subfigure}
		\begin{subfigure}[b]{0.49\textwidth}
		\includegraphics[width=\textwidth]{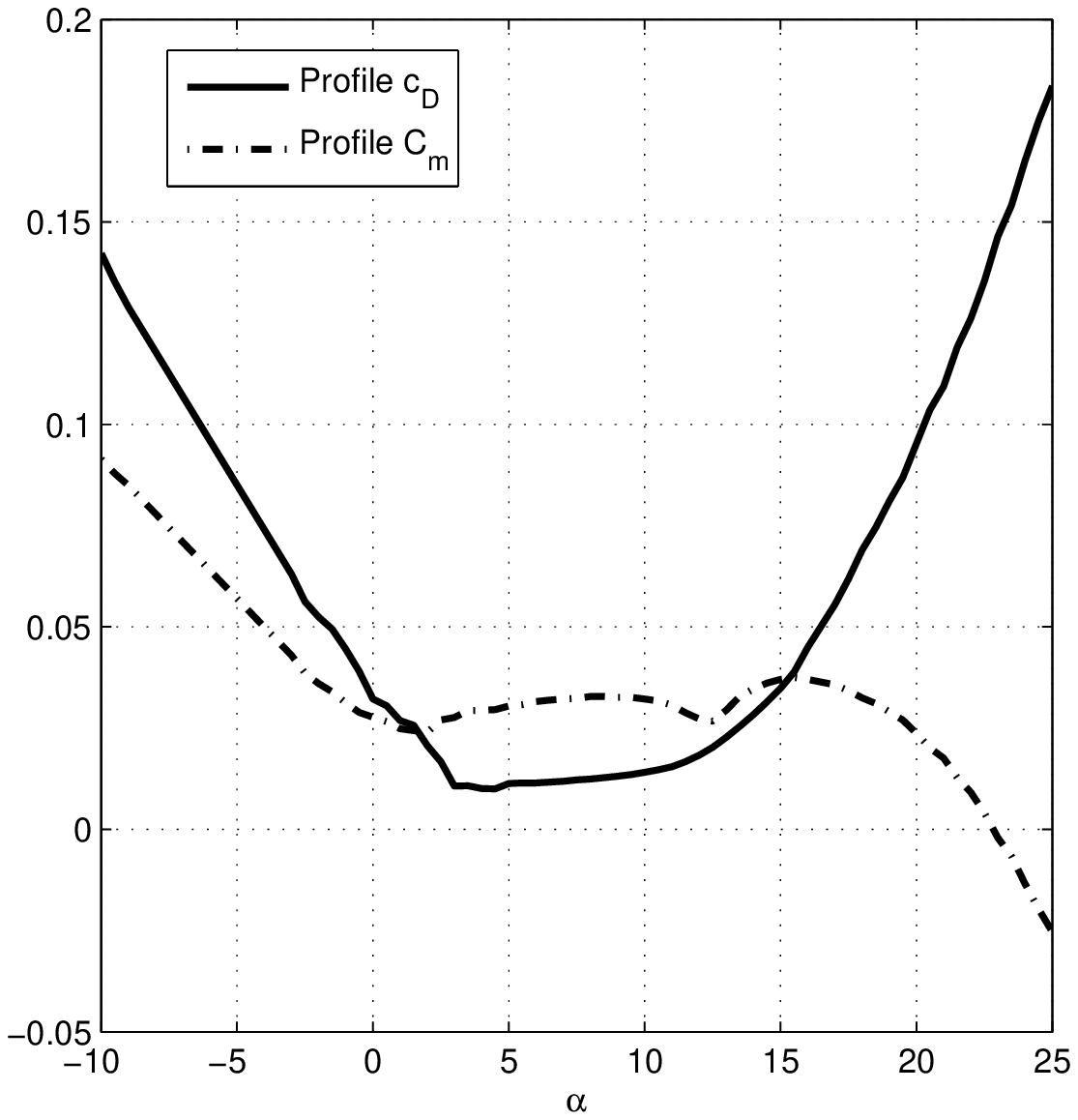}
		\label{fig:profDrag}
		\end{subfigure}
	\caption{Section data obtained with XFOIL.}
	\label{fig:secData}
\end{figure*}

Here, calculations of one flying wing are presented.
The wing is shown in Fig.\ref{fig:wing}, and is similar to delta-wing speed-kites.
Winglets and main wing are discretized by 24 and 100 Vortex elements.
A single airfoil along the wingspan is assumed, albeit with a different washout.
The 2D curves of lift, drag and moment were calculated using XFOIL and can be seen in Fig.\ref{fig:secData}.

\subsection{Local Lift and Coefficients}
\begin{figure*}[tbp]
	\centering
		\includegraphics[width=1.00\textwidth]{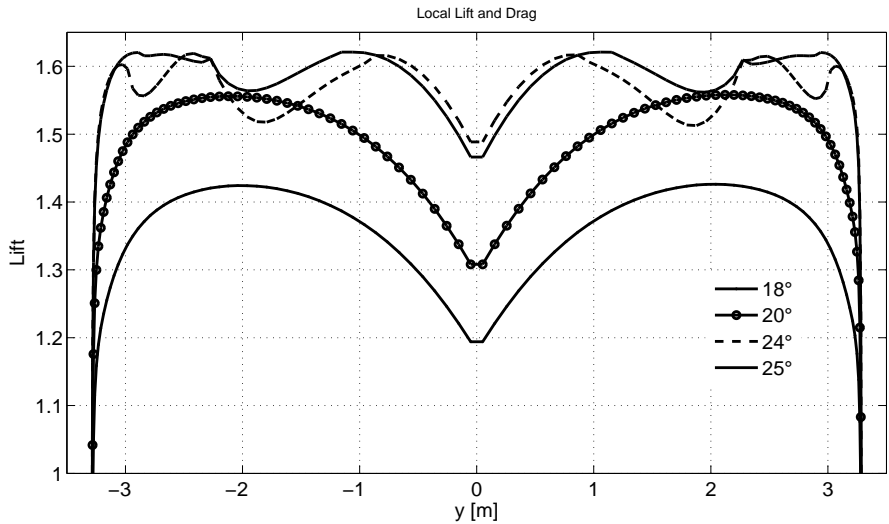}
	\caption{Local lift coefficient. Winglets not shown. Stalled regions can be seen at $24\textdegree$ and $25\textdegree$.}
	\label{fig:localLift}
\end{figure*}

\begin{figure*}[tbp]
	\centering
		\begin{subfigure}[b]{0.49\textwidth}
		\includegraphics[width=\textwidth]{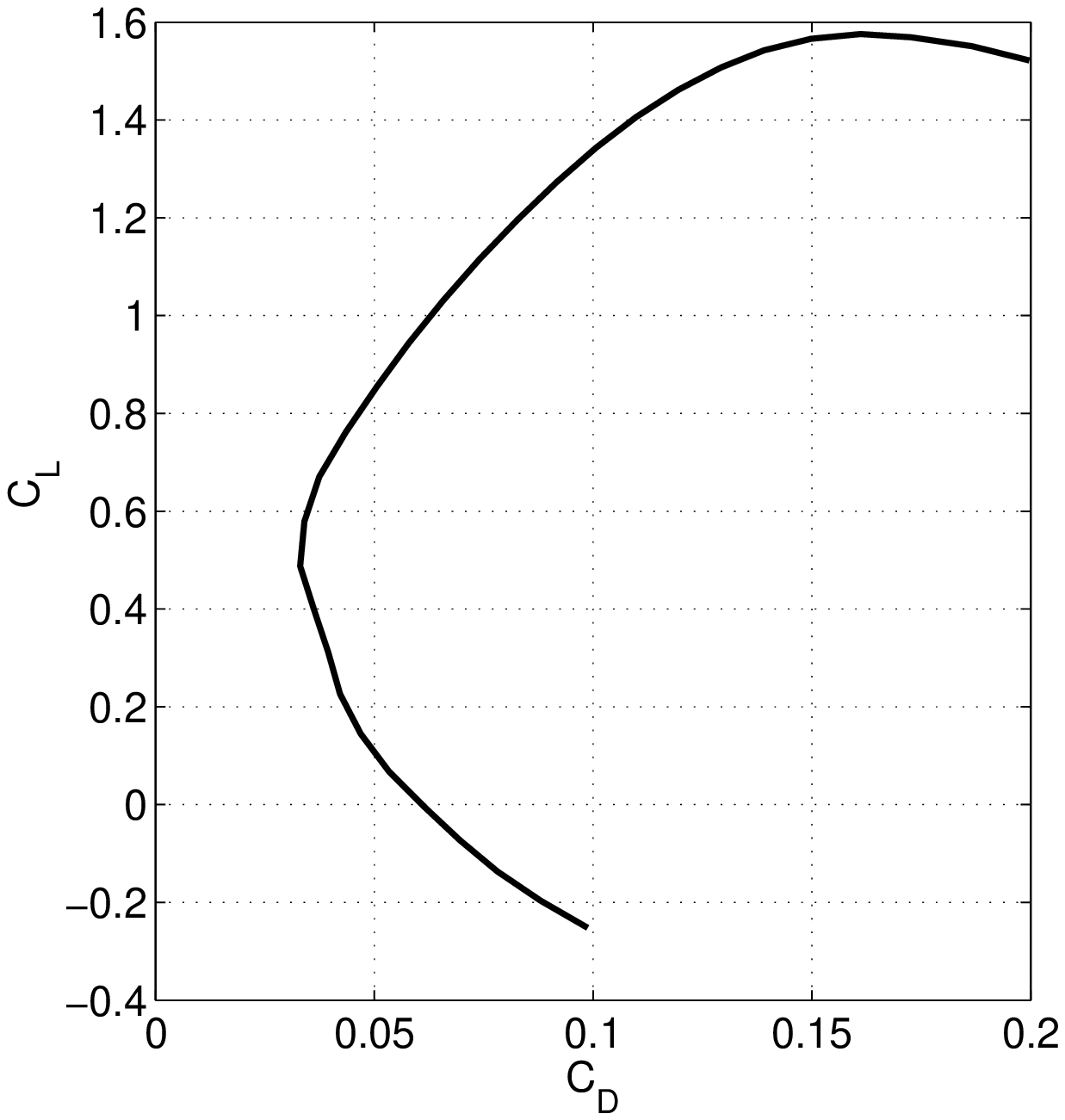}
		\label{fig:polardiagramm}
		\end{subfigure}
		\begin{subfigure}[b]{0.49\textwidth}
		\includegraphics[width=\textwidth]{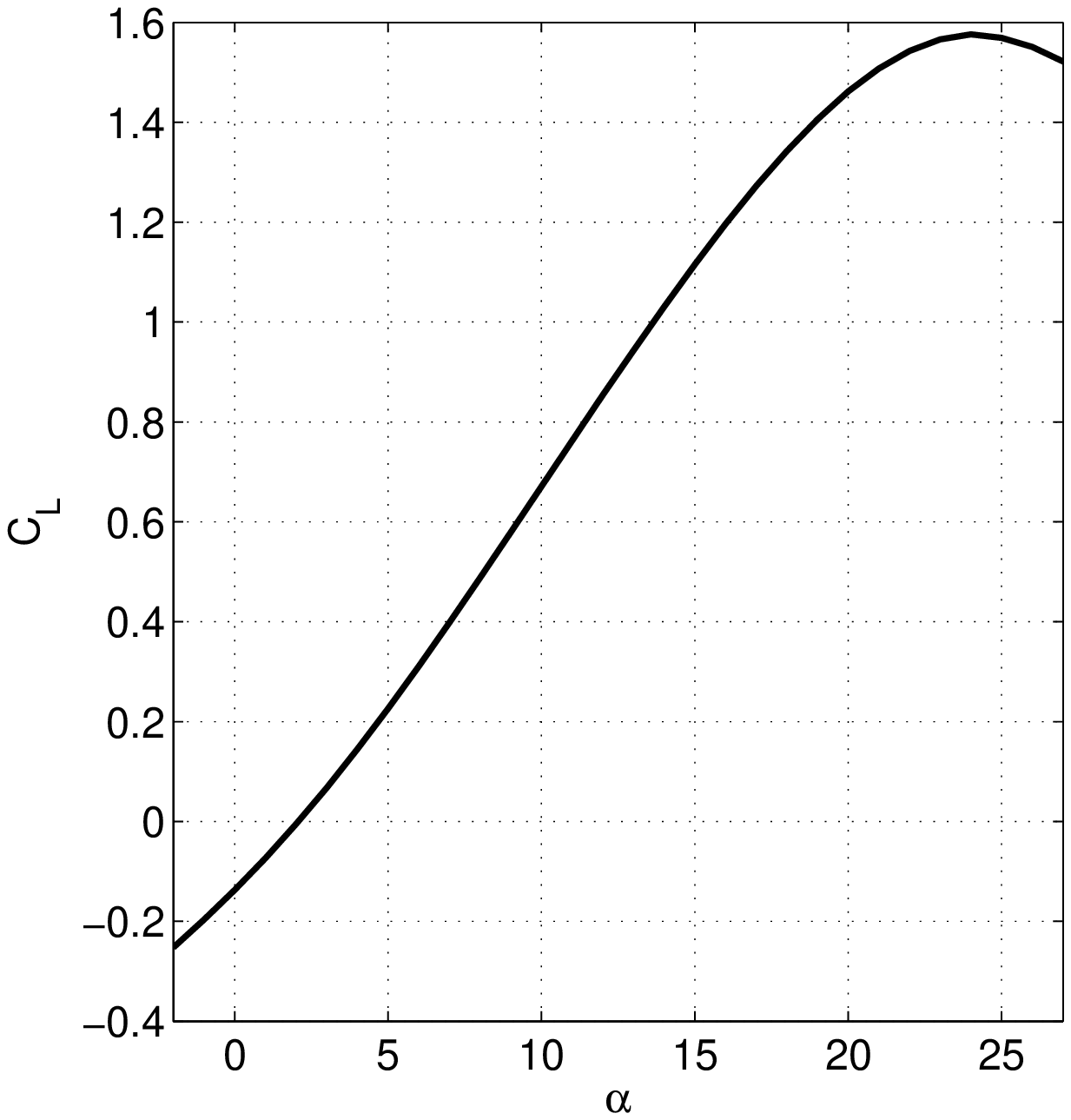}
		\label{fig:clkurve}
		\end{subfigure}
	\caption{Basic aerodynamic wing coefficients.}
	\label{fig:basicWing}
\end{figure*}

The local lift at zero sideslip at different angles of attack is shown in Fig.\ref{fig:localLift}.
Above $20\textdegree$ stall regions appear, where the local angle of attack is above the profile maximum.
The general coefficients are shown in Fig.\ref{fig:basicWing}.
Below and above the given angles of attack, no interpolation is possible and thus no results are available. 
The coefficient of maximum lift is a bit below 1.6, while the maximum section lift data is a bit above 1.6.
Next to the stalled region of the coefficient curve, it is also of interest that the lift slope of the aircraft is not linear within the unstalled region of low angles of attack.
This leads to different optimal angles of attack for best glide-ratio and best climb performance.

\subsection{Static Sideslip Stability}
\begin{figure*}[tbp]
	\centering
		\begin{subfigure}[b]{0.32\textwidth}
		\includegraphics[width=\textwidth]{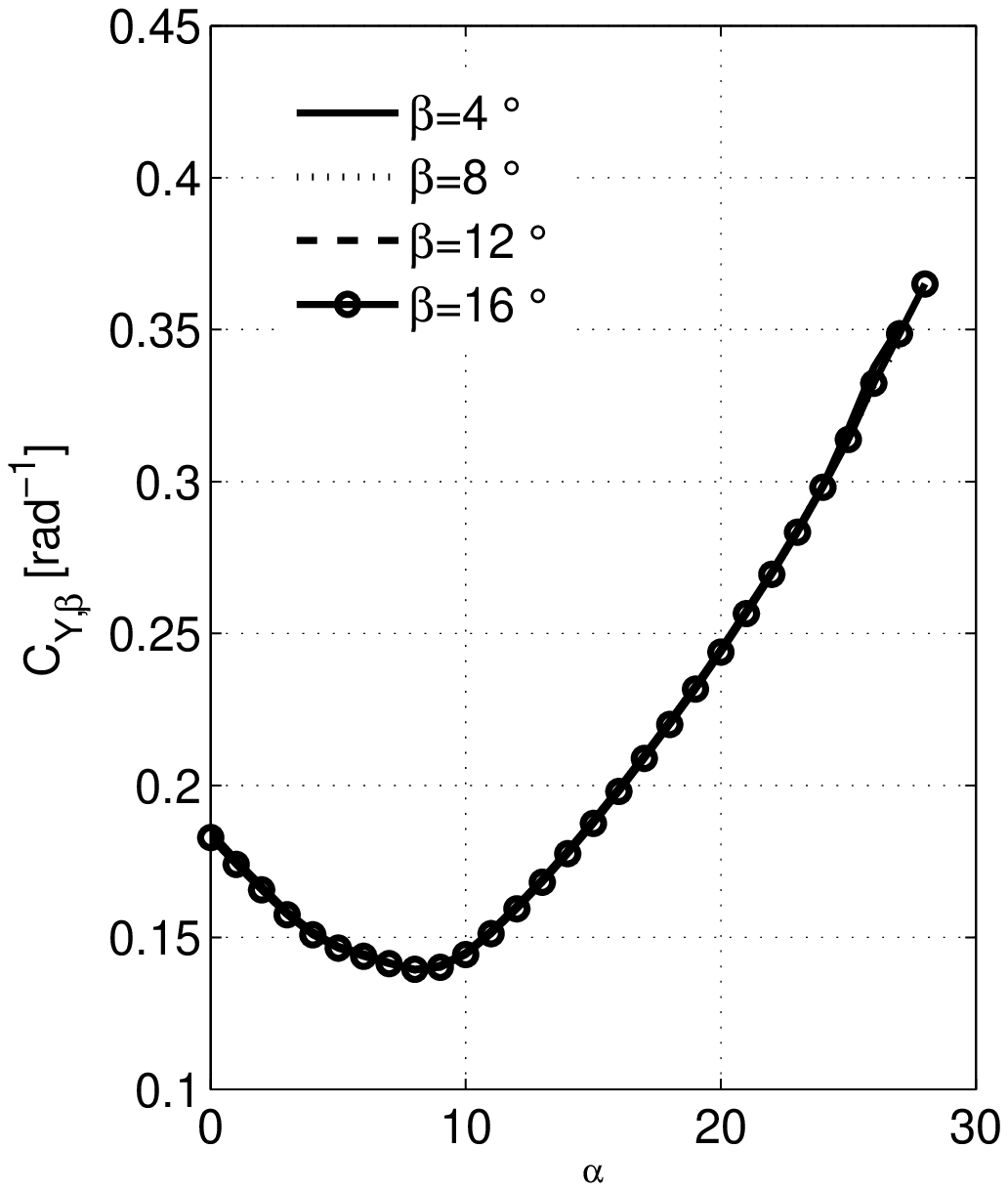}
		\end{subfigure}
		\begin{subfigure}[b]{0.32\textwidth}
		\includegraphics[width=\textwidth]{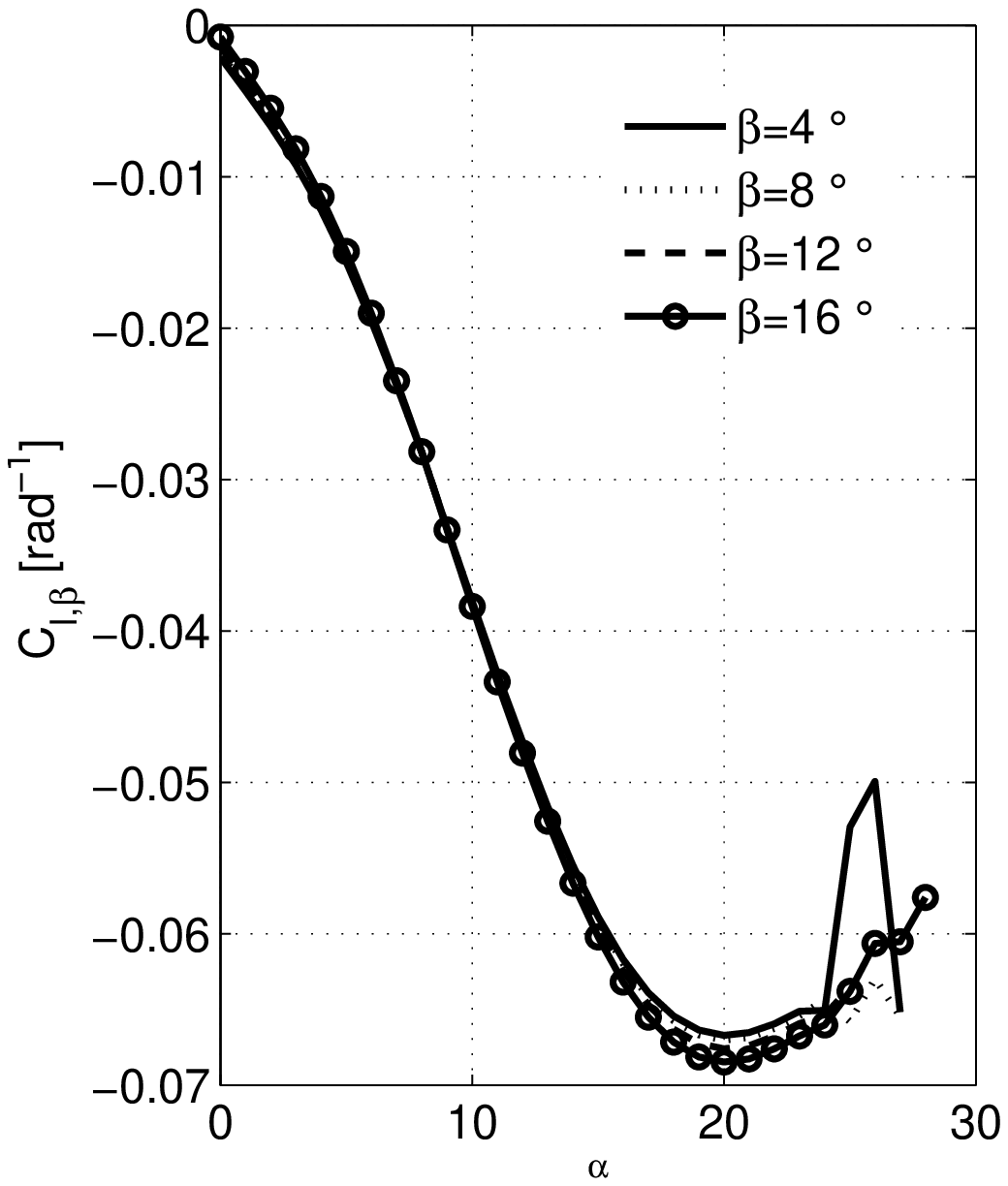}
		\end{subfigure}
		\begin{subfigure}[b]{0.32\textwidth}
		\includegraphics[width=\textwidth]{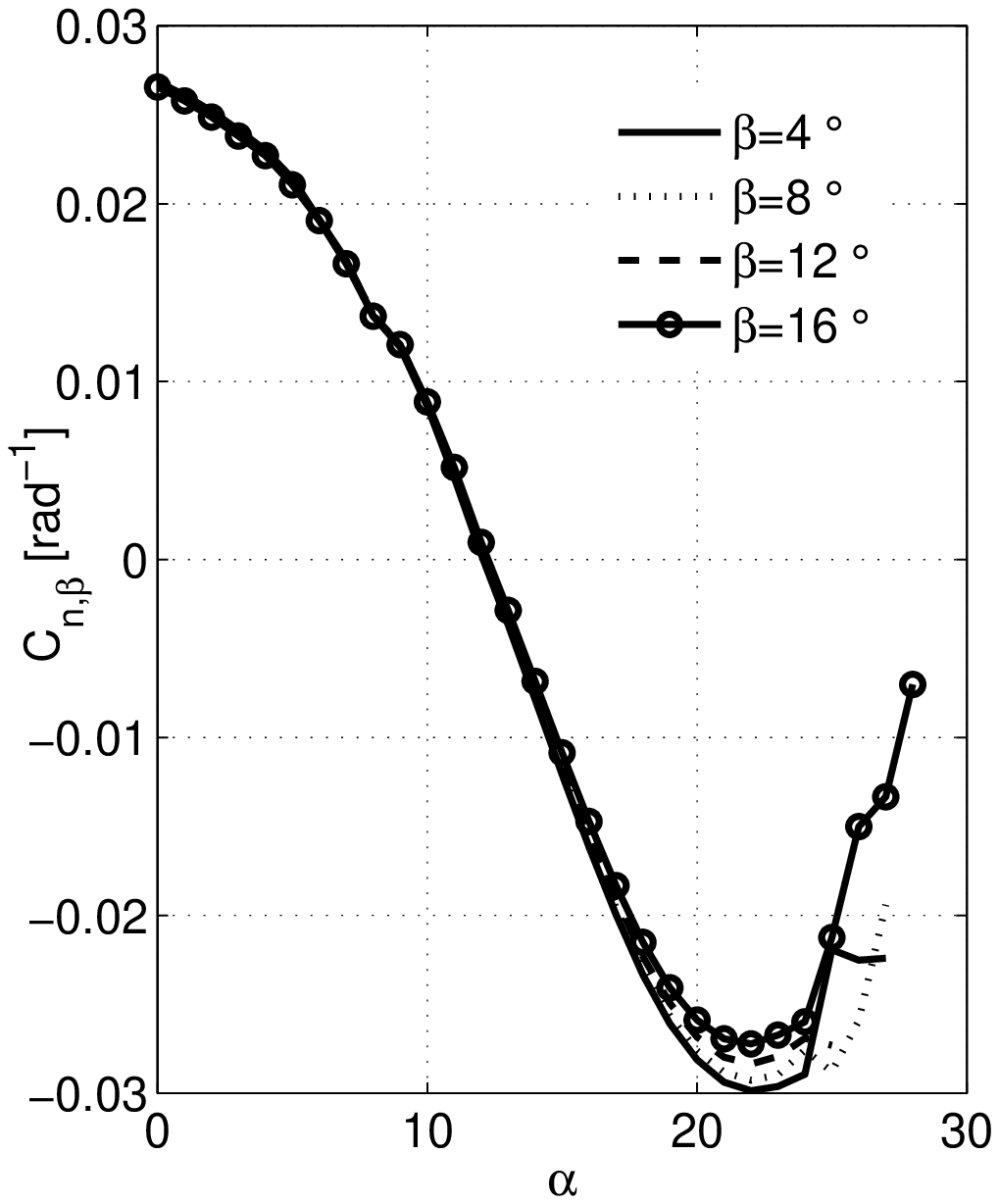}
		\end{subfigure}
	\caption{Lateral wing coefficients, evaluated at different sideslip angles.}
	\label{fig:lateralWing}
\end{figure*}

The sideslip results in a change in airspeed direction and thus in a change in lift and drag direction.
Additionally, the trailing vortex direction is changed.
All coefficients are meant in terms of body-coordinates.
That is, the sideforce points spanwise and the moments are meant in the aircraft coordinate system.
The point of reference is set to be the center of gravity which is quite close to the mean quarter-cord line.
The sideslip coefficients are the sideforce to to sideslip, the yawing moment due to sideslip and the rolling moment due to sideslip.
They can be seen in Fig.\ref{fig:lateralWing}, where the coefficients were evaluated at different sideslip angles.
However, the coefficients divided by the sideslip angle are in more or less the same for all sideslip angles and a linear coefficient for all sideslip angles can be assumed.
It can be seen that the directional stability decreases with the angle of attack and the aircraft becomes unstable above $12\textdegree$.
Lateral stability increases with angle of attack.

\section{Inverse Identification of Airfoil Section Data}
The results so far were concerned with the estimation of aerodynamic properties of 3-dimensional wings using 2-dimensional results from airfoil calculations.
Just as interesting is the problem of identifying the properties of the airfoil section using windtunnel measurements with 3-dimensional wings.
In fact this is quite common, even if the wing is simply a rectangle.
Instead of using endplates on rectangular models or using some formulas for downwash effects, the method presented here can be used to estimate the airfoil characteristics that lead to measurements with 3-dimensional models.
The results in this chapter were obtained with the following procedure.
Let $F_m = [CL_m, CD_m, CM_m]$ the coefficients of a three-dimensional wing as measured in a wind-tunnel.
The two-dimensional section characteristics are modeled with a linear function over the angle of attack with the unknown coefficients $x = [cl,cd,cm]$ and the resulting 3-dimensional characteristics from the nonlinear Vortex-Lattice method is $F_{3D}(x)$.
Then, the solution to the following problem should result in 2-dimensional airfoil characteristics that best describe the measurements under the assumption that the nonlinear Vortex-Lattice method is applicable here.
\begin{align*}
\min_{x} \left\| F_{3D}(x)-F_m\right\| 
\end{align*}
However, there are some problems with simply minimizing this function.
The method calculates apparent angles of attack at every section, which are based on the global lift distribution. 
This relationship is not necessarily unique.
Additionally, it can be ill-defined.
Due to downwash, the maximal lift coefficients of the airfoil data are only occurring in the center of a wing, and the local angles of attack are all shifted towards the zero-lift angle of attack.
For low aspect-ratio wings, this problem becomes more pronounced.
To guide the solution towards sensible scenarios, a regularization penalty is introduced.
Here, the second derivative of the airfoil data with respect to angle of attack is used which results in the following problem with the regularization parameter $\mu$
\begin{align*}
\min_{x} \left\| F_{3D}(x)-F_m\right\|^2 + \mu \left\| \pd[2]{}{\alpha} x \right\|^2.
\end{align*}
For linear coefficient functions, this results in no penalty.
The problem is solved by a Gauss-Newton Iteration, and the gradient of $F_{3D}$ with respect to $x$ is computed using finite differences.
The method has been applied to the identification of the Selig S1223\cite{S1223} profile from a rectangular wing with an aspect ratio of 2.7.
The airfoil exhibits highly nonlinear behavior and is prominent in low Reynolds-Number applications.
Good measurements for this airfoil have been obtained at UIUC\cite{SeligData}, with which the measurements and reconstructions will be compared.

\begin{figure*}[p]
	\centering
		\includegraphics[width=0.95\textwidth]{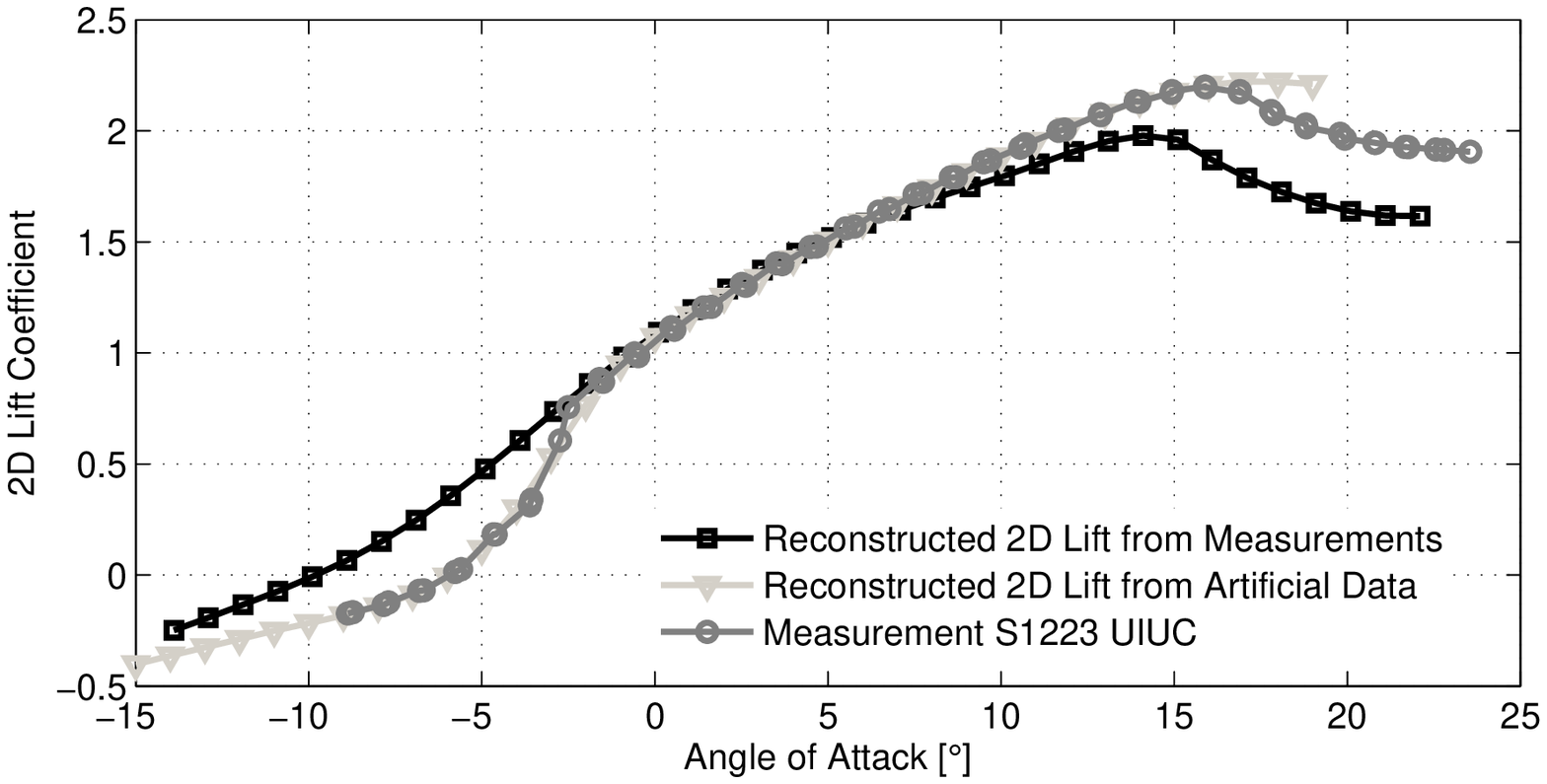}
	\caption{Comparison of Section Lift from the S1223 airfoil reconstructed from measurements at the windtunnel at the BIONIK Institute at the TU Berlin with Measurement Data from the UIUC\cite{SeligData}. AR 2.7, RE 200,000.}
	\label{fig:CLmeas2D}
\end{figure*}

\begin{figure*}[p]
	\centering
		\includegraphics[width=0.95\textwidth]{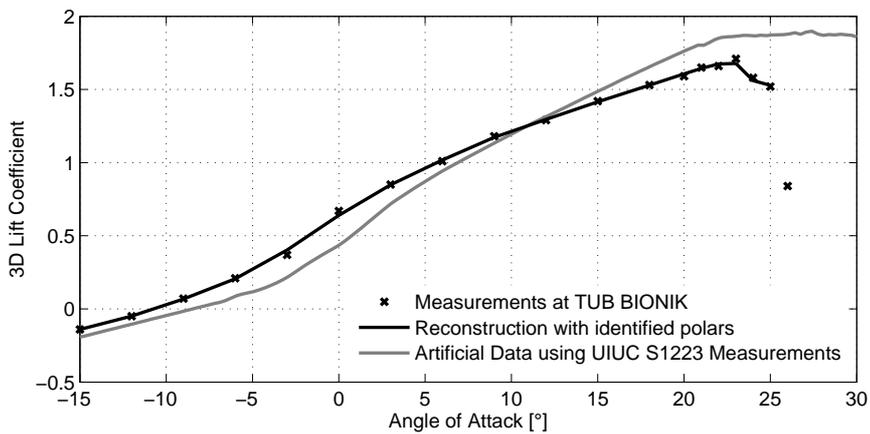}
	\caption{Comparison of Wing Lift with the S1223 airfoil at the windtunnel at the BIONIK Institute at the TU Berlin with measurements and forward calculation using Measurements from the UIUC\cite{SeligData}. AR 2.7, RE 200,000.}
	\label{fig:CLmeas3D}
\end{figure*}

The measurements were taken at a small windtunnel at the BIONIK Institute at the TU Berlin.
In Fig.\ref{fig:CLmeas3D} the measurements taken in Berlin, together with the artificial measurements obtained using the presented method with the reconstructed polars as well as the polars from the UIUC.
In Fig.\ref{fig:CLmeas2D} the 2D Lift coefficients reconstructed with the inverse method from the wing measurements and the artifical data can be seen, as well as the original S1223 measurements at UIUC.

At Lift coefficients between 1 and 1.5, the reconstruction agrees well with the measurements of the UIUC\cite{SeligData}.
At the Reynolds Number the maximum Lift coefficient should exceed 2.0, which is not reconstructed here.
Instead, the section lift seems to be stalling earlier.
Additionally, the characteristic drop in lift below 2\textdegree\ is not reconstructed but only hinted.

Using the artificial data, the 2D profiles are reconstructed very well until maximum lift, only the characteristic drop at around -2\textdegree\  has been slightly smoothed.
Hence, the inverse calculation works rather well.
However, using the measurements taken in Berlin, the airfoil data is not reconstructed well.
Either the wing has not been constructed close enough to the original S1223 specifications or the nonlinear Weissinger method is not applicable to these low aspect ratios.

\begin{thebibliography}{91.}%


\bibitem{XFOIL} Drela, Mark. "XFOIL: An analysis and design system for low Reynolds number airfoils." Low Reynolds number aerodynamics. Springer Berlin Heidelberg, 1989. 1-12.

\bibitem{3DDrela} Drela, Mark. "Three-dimensional integral boundary layer formulation for general configurations." 21st AIAA Computational Fluid Dynamics Conference. 2013.

\bibitem{Nonlinear47} Sivells, James C., and Robert H. Neely. "Method for calculating wing characteristics by lifting-line theory using nonlinear section lift data." No. NACA-TN-1269. National Advisory Committee for Aeronautics Langley Field VA Langley Aeronautical Laboratory, 1947.


\bibitem{Nonlinear76} Piszkin, Stanley T., and E. S. Levinsky. "Nonlinear lifting line theory for predicting stalling instabilities on wings of moderate aspect ratio." No. CASD-NSC-76-001. General Dynamics San Diego CA Convair Div, 1976.


\bibitem{NonlinearStability} Anderson, M. K. "Aerodynamic modeling for global stability analysis." AIAA Paper 4805 (2002): 2002.

\bibitem{VanDam} Van Dam, C. P. "The aerodynamic design of multi-element high-lift systems for transport airplanes." Progress in Aerospace Sciences 38.2 (2002): 101-144.


\bibitem{Decambering} Mukherjee, Rinku, Ashok Gopalarathnam, and Sung Wan Kim. "An iterative decambering approach for post-stall prediction of wing characteristics using known section data." AIAA Paper 1097 (2003): 2003.

\bibitem{Decambering2} Mukherjee, Rinku, and Ashok Gopalarathnam. "Poststall prediction of multiple-lifting-surface configurations using a decambering approach." Journal of aircraft 43.3 (2006): 660-668.

\bibitem{exDecambering} Sutton, Matthew. "A Superposition Approach for Determining Lift Distributions on Maneuvering Airplanes with Applications to Post Stall." (2010).

\bibitem{Weissinger} Weissinger, J. "Der schiebende Tragfl\"ugel bei gesunder Str\"omung." Jbuch Deutsch. Luftfahrtforschung, Flugwerk (1940).










\bibitem{Cart3D} Sturdza, Peter, et al. "A Quasi-Simultaneous Interactive Boundary-Layer Model for a Cartesian Euler Solver." AIAA paper for 50th AIAA Aerospace Sciences Meeting. 2012.

\bibitem{SeligData} Giguere, P., C. N. Ninham, and J. J. Guglielmo. "Summary of low-speed airfoil data. Vol. 1." Virginia Beach, VA: SoarTech Publications, 1995.

\bibitem{S1223} Selig, Michael S., and James J. Guglielmo. "High-lift low Reynolds number airfoil design." Journal of aircraft 34.1 (1997): 72-79.

\end{thebibliography}
\end{document}